\documentclass{EPS}

\usepackage{lineno}
\usepackage[dvipdfmx]{color}
\usepackage{comment}
\usepackage[dvipdfmx]{graphicx}
\usepackage{bm}
\usepackage{amsmath,amssymb}
\usepackage{natbib}

\title{Orbital Evolution of a Circumbinary Planet in a Gaseous Disk}
\author{
Akihiro Yamanaka, Department of Astronomy, Kyoto University, Kitashirakawa-Oiwake-cho, Sakyo-ku, Kyoto, 606-8502, Japan, akihiro@kusastro.kyoto-u.ac.jp\\
Takanori Sasaki, Department of Astronomy, Kyoto University, Kitashirakawa-Oiwake-cho, Sakyo-ku, Kyoto, 606-8502, Japan
}

\abstract{
Sub-Jupiter classed circumbinary planets discovered in close-in binary systems have orbits just beyond the dynamically unstable region, which is determined by the eccentricity and mass ratio of the host binary stars.
These planets are assumed to have formed beyond the snow line and migrated to the current orbits rather than forming {\it in situ}. We propose a scenario in which a planet formed beyond the snow line and migrated to the inner edge of the circumbinary disk, which was within the unstable area, and then moved to the current orbit through outward transportation. This outward transportation is driven by the balance of orbital excitation of the central stars inside the gravitationally unstable region and damping by the gas-drag force. We carried out $N$-body simulations with a dissipating circumbinary protoplanetary disk for binary systems with different eccentricities and mass ratios. Planets are more likely to achieve a stable orbit just beyond the unstable region in less eccentric binary systems. This result is not as sensitive to mass ratio as it is to eccentricity. These dependencies are consistent with the data from observed binary systems hosting circumbinary planets.
We find CBPs' orbits close to the instability boundaries are explained by our orbital evolution scenario.
}


\begin{document}

\maketitle

\section{Introduction}

The {\it Kepler} satellite has discovered many close-in circumbinary planets (CBPs), namely
Kepler-16b \citep{kepler16b}, Kepler-34b \citep{kepler34and35b}, Kepler-35b \citep{kepler34and35b}, Kepler-38b \citep{kepler38b}, Kepler-47b, Kepler-47c \citep{kepler47bc,kepler47d}, Kepler-47d \citep{kepler47d} PH1(Kepler-64b) \citep{ph1}, Kepler-413b \citep{kepler413b}, Kepler-453b \citep{kepler453b}, and Kepler-1647b \citep{kepler1647b}.
Table \ref{tab:obsdata} shows the mass, semi-major axis and orbital eccentricity of the circumbinary planetary systems observed by the {\it Kepler}.
The spectral types of the binary host stars are KM (Kepler-16, -413), GG (Kepler-34, -35, -1647), GM (Kepler-38), and FM (Kepler-64).
Their masses are $\sim 1M_{\odot}$, the primary and secondary stars are separated by $\sim 0.2$ au, and the orbital eccentricities are $\sim 0.1$ except for the Kepler-34 and PH1 (Kepler-64Aa and Kepler-64Ab) systems, which have high eccentricities of 0.521 and 0.212, respectively. Most of these planets have close-in orbits with semi-major axes under 1 au, and they have Neptune-class mass (except for the Kepler-47 and Kepler-453 systems). Another characteristic is that only the Kepler-47 system hosts three CBPs, whereas the other binary systems host only one planet \citep{kepler47bc}.

\begin{table}
  \centering
  \begin{tabular}{c|ccccccc}
    \hline
    & $M_\mathrm{p} [M_{\odot}]$ & $M_\mathrm{s} [M_{\odot}]$ & $e_{\mathrm{bin}}$ & $a_{\mathrm{bin}}$ [au] & $M_{\mathrm{pl}} [M_{\mathrm{J}}]$ & $a_{\mathrm{pl}}$ [au] & $e_{\mathrm{pl}}$ \\
  \hline
    Kepler-16b & 0.687 & 0.202 & 0.16 & 0.224 & 0.333 & 0.72 & 0.0069 \\
    Kepler-34b & 1.049 & 1.022 & 0.521 & 0.228 & 0.22 & 1.086 & 0.182 \\
    Kepler-35b & 0.885 & 0.808 & 0.142 & 0.176 & 0.127 & 0.605 & 0.042 \\
    Kepler-38b & 0.949 & 0.249 & 0.103 & 0.147 & 0.384 & 0.464 & 0.032 \\
    Kepler-47b & 1.043 & 0.362 & 0.023 & 0.084 & 0.0065 & 0.288 & 0.021 \\
    Kepler-47d & - & - & - & - & 0.060 & 0.699 & 0.024 \\
    Kepler-47c & - & - & - & - & 0.010 & 0.964 & 0.044 \\
    PH1 & 1.53 & 0.378 & 0.212 & 0.174 & 0.531 & 0.634 & 0.0702 \\
    Kepler-413b & 0.82 & 0.542 & 0.037 & 0.099 & 0.211 & 0.355 & 0.07 \\
    Kepler-453b & 0.944 & 0.195 & 0.0524 & 0.185 & $6.29 \times 10^{-4}$ & 0.79 & 0.118 \\
    Kepler-1647b & 1.2207 & 0.9678 & 0.1602 & 0.1276 & 1.52 & 2.72 & 0.0581 \\
  \hline
  \end{tabular}
  \caption{Parameters of close-in CBPs discovered by Kepler. Index p, s, bin, and pl represent primary, secondary, binary, and planet, respectively. For reference, please see the main text.}
  \label{tab:obsdata}
\end{table}

The gravitational potential around the binary system varies as binary stars rotating around the common center of gravity influence the planet-forming environment in the binary system. When a planet enters within a certain semi-major axis, its orbit experiences strong excitation by the time-varying gravitational potential, leading to the orbital instability.
Numerous works have been done on this topic \citep[e.g.][]{dvorak1984,crit1,DB2011,KS2014,LK2018}.
In \cite{Quarles2018},
the sizes of this unstable region for co-planar orbits were
derived by $N$-body simulations as a function of the binary eccentricity and mass ratio,

\begin{linenomath} \begin{equation}
  r_{\mathrm{crit}} = (1.48 + 3.92e_{\mathrm{bin}} - 1.41e_{\mathrm{bin}}^2 + 5.14\mu + 0.33e_{\mathrm{bin}}\mu - 7.95\mu^2 + 4.89e_{\mathrm{bin}}^2\mu^2)a_{\mathrm{bin}},
  \label{eq:crit_Q2018}
\end{equation} \end{linenomath}
where $e_{\mathrm{bin}}, a_{\mathrm{bin}}$, and $\mu$ are the eccentricity, separation, and mass ratio (defined as $M_{\mathrm{s}}/(M_{\mathrm{p}}+M_{\mathrm{s}})$) of the binary stars, respectively.

\begin{figure}[p]
  \centering
  \includegraphics[width=12.0cm]{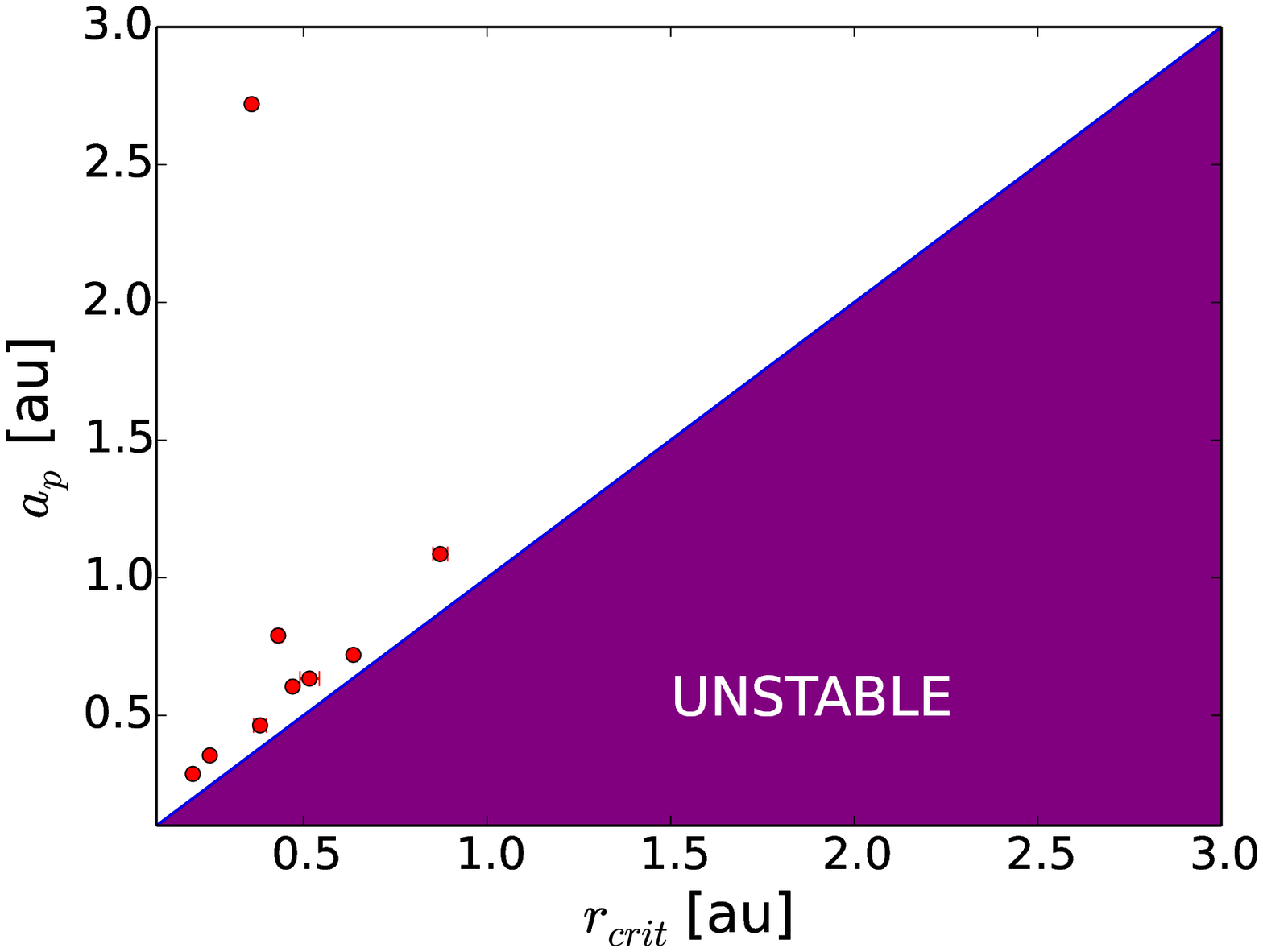}
  \caption{Semi-major axis $a_{\mathrm{p}}$ and critical radius $r_{\mathrm{crit}}$ calculated using equation (\ref{eq:crit_Q2018}) \citep{Quarles2018} in observed circumbinary planet systems (except for Kepler-47c, and Kepler-47d, which are not the innermost planets of the system). The solid blue line represents $a_{\mathrm{p}} = r_{\mathrm{crit}}$ and the area under this line, which is colored in purple, represents dynamically unstable area. Most of the observed planets' semi-major axes are lined up just above the critical radius.\label{a_vs_crit}}
\end{figure}

By comparing this critical radius with the observed semi-major axis for systems except Kepler-47c, -47d, and -1647b, one can find that the observed planets' orbits line up just over the boundary of the unstable regions (Figure \ref{a_vs_crit}).
Observed CBPs are expected to be formed in the outer region of the protoplanetary disk and then migrate to the current orbit.
Because of the large sizes of the observed CBPs, which are mostly $0.1-0.3 M_{\mathrm{J}}$, they would form beyond the snow lines, where the increased dust surface-density enables the CBPs to grow to sub-Jupiter size. Then, CBPs would migrate to the current orbits.

Several hydrodynamic simulations of observed binary systems have recently been carried out \citep[e.g.][]{TK2018,PN2013}
with aim of reproducing such CBPs' orbits.
Numerical simulations were conducted on the Kepler-16, -34, -35, -38, and -413 systems in \cite{TK2018}. Their results implied that the location of the disk's inner edge would be farther out than previously estimated by \cite{cavity}. They also showed that the migration of planets embedded in the outer area of the protoplanetary disk halted at the disk edge,
which was far beyond the observed current orbits.
\cite{PN2013} conducted hydrodynamic simulations of the circumbinary-disk structures around the Kepler-16, 34, and 35 systems and the results showed that the disk-surface density peaks at approximately $R_{\mathrm{peak}}\sim (3.8+35e_{\mathrm{d}}) a_{\mathrm{bin}}$ where $e_{\mathrm{d}}$ is disk eccentricity. They also simulated type-I protoplanet migration in a circumbinary disk whose surface-density profile and eccentricity are those obtained using hydrodynamic simulations. The inward migration of protoplanets from the outer region stops at the disk's surface-density peak, which is far distant from the current orbit in each system.
These results show that planet migration would halt far beyond the observed current orbits. Mechanisms to shrink the disk's inner cavity and let the planet migrate
at least to the current orbit
are required to reproduce the present orbit of the observed CBPs; however, such mechanisms have not been yet revealed.

\begin{figure}[htbp]
  \centering
  \includegraphics[width=12.0cm]{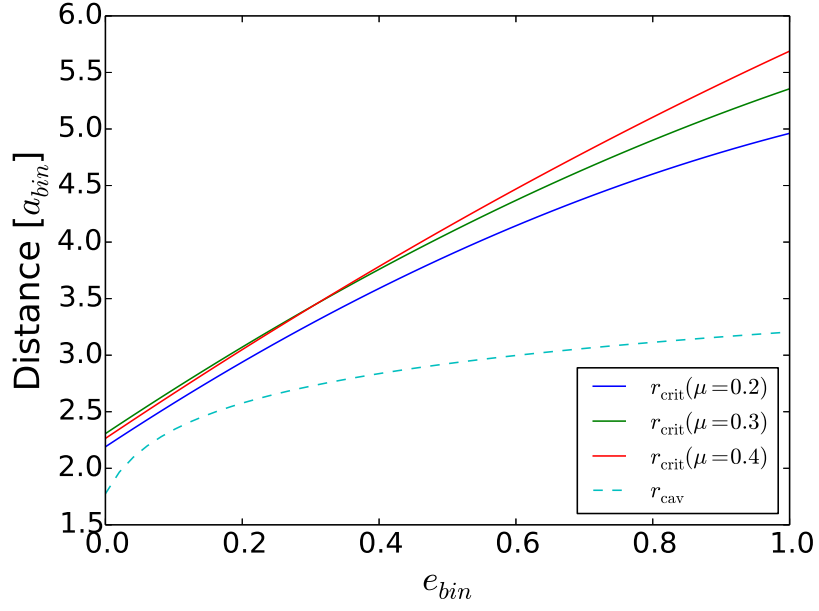}
  \caption{Solid lines represent the locations of the instability boundary for $\mu= 0.2, 0.3, 0.4$ binary systems. The dashed line represents the inner cavity size and critical radius for a $\mu=0.3$ binary system. The location of unstable boundary is always larger than the disk inner edge for any $e_{\mathrm{bin}}$. $r_{\mathrm{crit}}$ and $r_{\mathrm{cav}}$ are calculated using equation \ref{eq:crit_Q2018} and equation, \ref{eq:cav} respectively.}
  \label{fig:cav_crit}
\end{figure}

On the other hand, using SPH simulations, \cite{cavity} showed that the inner edge of the circumbinary disks are truncated close to the 3:1 resonance, which is $r = 2.08a_{\mathrm{bin}}$, for nearly circular binaries and close to the 4:1 resonance, $r = 2.52a_{\mathrm{bin}}$, for binaries with $e_{\mathrm{bin}}>0$. While the size of the truncation has a strong dependence upon binary eccentricity, its dependency on the binary-mass ratio $\mu$ is small. The inner disk edge for $\mu = 0.3$ is derived as
\begin{linenomath} \begin{equation}
  r_{\mathrm{cav}} = (0.425\ln (e_{\mathrm{bin}} + 0.0358) + 3.19)a_{\mathrm{bin}}.
  \label{eq:cav}
\end{equation} \end{linenomath}
Figure \ref{fig:cav_crit} plots the semi-major axis of the inner cavity and the critical radius (Equation \ref{eq:crit_Q2018}) for the $\mu=0.2, 0.3, 0.4$ binary system as functions of $e_{\mathrm{bin}}$. The disk inner edge (dashed line) is inside of the locations of the instability boundary (solid lines) for all $e_{\mathrm{bin}}$.
Although the main physics of truncation of the gas disk is the gravity of central binary, which is also responsible for the orbital instability of the CBP, viscous spreading allows the disk-gas to spread inside the unstable area.

\cite{cavity} indicated that the location of the inner edge was determined by Lindblad torques, and we can take this radius $r_{\mathrm{cav}}$ as the lower-limit of the inner cavity radius. So, if the disk's inner edge somehow shrank to the lower limit radius, the disk inner edge could have once been located inside of the critical radius. In this case, CBPs could also have migrated near or inside of the critical radius, so that we should consider the gravitational stability of the planets near the unstable boundaries.

The orbital evolution and gas accretion of a protoplanet embedded in a circumbinary disk has been studied in \cite{giant_in_gas}. They performed a hydrodynamic simulation of a circumbinary disk until the system reached a quasi-equilibrium state and the protoplanet's orbital evolution was solved using the obtained disk profiles for different planetary masses. They showed that
a Saturn-mass planet embedded at the disk's inner edge first migrates inward, and soon, migration reversal occurs by increase of planet eccentricity before it reaches the 4:1 resonance. Then, the planet migrates outward and its orbit remains stable.
On the other hand, a planet more massive than Jupiter, which migrates faster than a Saturn-mass planet, reaches and is trapped into 4:1 resonance with the binary before the occurrence of migration reversal. The resonance rapidly increases the planet's eccentricity until it undergoes a close encounter with the central binary, leading to planet ejection.
Their works offer a hint to maintaining a stable orbit near the unstable boundary via orbital evolution from the inside the unstable area.

In this study we consider a case where the inner boundary of the disk is located in a gravitationally unstable area and the planet enters the unstable region via type-I migration. We examine whether a stable CBP orbit near the unstable boundary can be achieved via outward transportation from the unstable area and search for the ranges of binary eccentricity and mass ratio in which a planet can maintain a stable orbit. Present orbits of CBPs clearly show they have experienced inward migration at least to current distance, although how the planets migrate inside the unstable area is yet unclear. Our aim is to demonstrate whether a planet can maintain a stable orbit just outside the unstable boundary even if the planet once entered unstable area via inward migration. The detailed process of inward migration of a planet and complex feedback between the disk and planet are not considered in this study.

\section{Our Scenario}

\begin{figure}
  \centering
  \includegraphics[width=14.0cm]{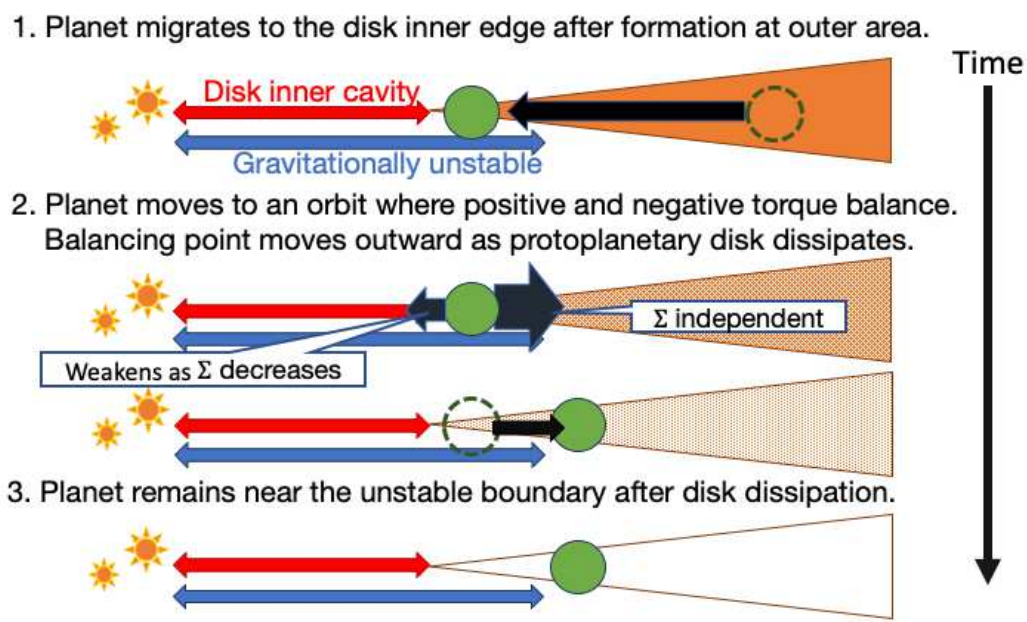}
  \caption{Orbital-evolution scenario. 1: Protoplanet formed in the outer area migrates to the disk's inner edge, which is within $r_{\rm crit}$. 2: Planet moves to an orbit where positive and negative torques balance-out. This balancing point moves outward as the protoplanetary disk dissipates. 3: Planet remains near the unstable boundary after disk dissipation.\label{fig:scenario}}
\end{figure}

We assume that the disk's inner edge forms within the gravitationally unstable area in a binary system and suppose that a planet formed in the outer region in a protoplanetary disk migrates into the unstable area. Here, we consider that the inward migration of the planet halts at the disk's inner edge.  \cite{SF2015} showed $\sim 80 \%$ of destabilized CBPs were ejected from the gas-free systems, suggesting that destabilized CBPs commonly move outward from the unstable regions. So, when the planet experiences a severe gravitational perturbation in the unstable region, its orbit tends to become larger. We speculate, in a gaseous disk with high enough gas surface density, the drag force from the gas can damp the orbital excitation and thus the planet can sustain a stable orbit at the distance where damping by the gas-drag force counterbalances the orbital excitation by the host binary. The excitation of the orbit is determined by the gravitational interaction between the planet and the host binary and thus does not depend upon the disk-gas density, while the gas-drag force becomes weak as the gas density decreases by dissipation. As the gas-drag force weakens, the semi-major axis at which excitation and damping balance-out becomes larger, resulting in the outward transportation of the planet. We examine the possibility of a planet emigrating from the unstable region while enough gas remains to prevent it from escaping the system. (See Figure \ref{fig:scenario}.)

\subsection{Orbital Integration} \label{subsec:integration}

The orbital evolution of a planet is calculated using the $N$-body calculation with the fourth-order Hermite method. The equation of motion is given as follows

\begin{linenomath} \begin{equation}
  \bm{a}_i = \begin{cases}
    -\sum_{j\neq i}\frac{GM_j}{r_{ij}^3}(\bm{r}_i - \bm{r}_j) & (r_i < r_{\mathrm{cav}}) \\
    -\sum_{j\neq i}\frac{GM_j}{r_{ij}^3}(\bm{r}_i - \bm{r}_j) + f_{\mathrm{GD}}\delta_{i3} & (r_i \geq r_{\mathrm{cav}}),
  \end{cases}
  \label{EOM}
\end{equation} \end{linenomath}
where $r_{ij} = |\bm{r}_i-\bm{r}_j|$ and $f_{\mathrm{GD}}$ are
the acceleration induced by gas-drag force
and indices $i,j=1,2,3$ corresponds to the primary, secondary, and planet, respectively. The system's length and mass are normalized with the separation and total mass of the central binary. We adopt $0.2\mathrm{au}, 1M_{\odot}$ as typical values of binary separation and total mass, respectively.
And the mass of the planet is set to $0.1M_{\mathrm{J}}$.
Time normalization is $T_o=\sqrt{a_{\mathrm{bin}}^3/GM_{\mathrm{tot}}}$, where $a_{\mathrm{bin}} and M_{\mathrm{tot}}=M_{\mathrm{p}}+M_{\mathrm{s}}$ are the separation and total mass of the binary stars, respectively. In reality, circumbinary disks have nonaxisymmetric structures due to the gravitational field of the central binary stars.
In this study, however, in order to investigate the influence of presence of disk gas on planet-orbital evolution, we consider only the case that the gas-density profile to be axisymmetric and proportional to the minimum-mass solar nebula (MMSN) for a circumbinary protoplanetary disk beyond the inner cavity. This simplification serves to allow the investigation of the influence of the gaseous disk upon planet-orbital evolution.
Considering complex feedback between the disk and the planet is beyond our scope of this study.
The calculation is halted when the planet's orbital eccentricity exceeds 1, which we interpret that the planet has become gravitationally unbound.
The integration time-step is set to $2^{-8} P_{\mathrm{bin}}$.

The center of the system is at the center of the gravity of the binary. Although the exact location of the barycenter depends on the location of the planet, the offset from the barycenter of binary is negligible because the planet mass is smaller than the binary mass by three orders of magnitude.
A planet in the unstable region becomes unbound without a gaseous disk within several thousand orbital periods of the binary, which corresponds to $10 \sim 10^2$yr. This suggests that the reaction timescale of the planet to the gravitational perturbation from the central binary is shorter than typical timescale of gas dissipation($\sim 10^7$yr) by 5 orders of magnitude. Thus, the evolution of the planet's orbital and disk-surface density can be calculated independently. In this study, we assume the disk dissipates constantly across the gas disk. Disk-surface density is described using the disk parameter $f_{\mathrm{disk}}=\Sigma/\Sigma_{\mathrm{MMSN}}$, where $\Sigma_{\mathrm{MMSN}}$ represents surface density of a MMSN. Variation of $f_{\mathrm{disk}}$ is 1, 0.56, 0.316, 0.178, 0.10, and 0.05 following exponential dissipation of the protoplanetary disk. The first run is calculated with a given gas density to obtain the final semi-major axis and eccentricity of the planet. The next run is calculated with a slightly thinner gas density and the final orbital distance of the previous run. Disk-density evolution is modeled by repeating this process, and the motion of the planet is integrated for $10^5T_o$ in each run.

\subsection{Gas Drag} \label{subsec:gasdrag}

A planet in a gaseous disk experiences several kinds of gas-drag forces. In this study, we take $M_{\mathrm{pl}}=0.1M_{\mathrm{J}}=1.89\times 10^{29}\mathrm{g}$ to be typical planetary masses. When the object's mass is larger than $\sim 10^{25}\mathrm{g}$, the gravitational gas drag dominates damping of the object's orbit(\cite{gasdrag}, \cite{gasdrag2}). By using the damping timescale of gravitational gas drag $\tau_{\mathrm{grav}}$,
the acceleration by gas-drag is written as
\begin{linenomath} \begin{equation}
  f_{\mathrm{GD}} = - \frac{\bm{v} - \bm{v}_{\mathrm{gas}}}{\tau_{\mathrm{grav}}},
\end{equation} \end{linenomath}
where $\bm{v}, \bm{v}_{\mathrm{gas}}$ are the velocities of the planet and the gas, respectively. The damping timescale, $\tau_{\mathrm{grav}}$, is
\begin{linenomath} \begin{equation}
  \tau_{\mathrm{grav}} \simeq \left(\frac{M_{\odot}}{M} \right)\left(\frac{M_{\odot}}{\Sigma_{\mathrm{gas}}r^2} \right)\left(\frac{c_s}{v_{\mathrm{K}}} \right)^4 \Omega_{\mathrm{K}}^{-1},
  \label{eq:tau_grav}
\end{equation} \end{linenomath}
where $c_s$ and $v_{\mathrm{K}}$ are the sound velocity and the Keplerian velocity. It is worth noting the planet velocity $\bm{v}$ is not necessarily equal to the Keplerian velocity $v_{\mathrm{K}}$ because the planet experiences complex gravitational forces from binary and drag force from gaseous disk. Using the gas surface density of the MMSN, $\Sigma_{\mathrm{MMSN}}=1700(r/1\mathrm{AU})^{-3/2}\mathrm{gcm^{-2}}$, Equation (\ref{eq:tau_grav}) becomes
\begin{linenomath} \begin{equation}
  \tau_{\mathrm{grav}} \sim 3.9 \left(\frac{M_{\mathrm{p}}}{0.1M_{\mathrm{J}}}\right)^{-1}\left(\frac{\Sigma_{\mathrm{gas}}}{\Sigma_{\mathrm{MMSN}}}\right)^{-1}\left(\frac{r}{0.5\mathrm{AU}}\right)^2 \mathrm{year.}
\end{equation} \end{linenomath}
Supposing $\bm{v_{\mathrm{gas}}}=(1-\eta)\bm{v_{\mathrm{K}}}$ the gas-drag term is
\begin{linenomath} \begin{equation}
  f_{\mathrm{GD}} \propto M_{\mathrm{p}}f_{\mathrm{disk}}r^{-3/2},
  \label{eq:gasdrag}
\end{equation} \end{linenomath}
where $f_{\mathrm{disk}}$ is the disk-density parameter $\Sigma_{\mathrm{gas}}/\Sigma_{\mathrm{MMSN}}$.
$\eta$ is a dimensionless number that gives the velocity gap between the planet and the disk-gas, which is defined as $\eta = - (h/2r)^2\frac{\partial \ln{P}}{\partial \ln{r}}$, where $P$ and $h$ are gas pressure and scale height of the disk. According to equation (\ref{eq:gasdrag}), more massive planet experience stronger gas drag, but it is the ratio of gas drag to gravitational perturbation that determines the planet's orbital evolution. Gravitational perturbation on the planet is given numerically by calculating the gravitational force between the binary stars and the planet.

\subsection{Initial Conditions} \label{subsec:init}

We consider the planet is co-planar with the central binary in a two-dimension system.
Assuming that the planet migrates to the inner edge of the protoplanetary disk, it is initially placed at a distance $r_{\mathrm{cav}}$ from the center of gravity with a random azimuthal angle.
The initial velocity of the planet is set to the Kepler velocity, and the initial eccentricity is set $e_{0}=0$.
Although the truncation radius of the disk depends on the binary mass ratio, we adopt a value of $r_{\mathrm{cav}}$ (equation \ref{eq:cav}) for all $\mu$ because the dependence on $\mu$ is weak compared to that on $e_{\mathrm{bin}}$.
Note that although the actual planet's velocity would be between the Kepler velocity and the gas' velocity because the planet is dragged by the disk gas, the planet's initial velocity is set to the Kepler velocity for simplicity. This simplification does not have major effects on the outcomes
because the planet's velocity is modified immediately via gravitational interactions between the stellar binary and the drag force from the disk.
We investigate the planet's orbital evolution around binary stars with different mass ratios ($\mu = M_{\mathrm{s}}/(M_{\mathrm{p}}+M_{\mathrm{s}})$) and eccentricities ($e_{\mathrm{bin}}$). The planetary mass is set to $0.1M_{\mathrm{J}}$ and the initial gas-surface density is $1\Sigma_{\mathrm{MMSN}}$. The orbital evolution is calculated around binary systems with $e_{\mathrm{bin}}=0.05, 0.075, 0.10, 0.20$ and $\mu = 0.2, 0.3, 0.4$. We also conducted a series of calculations with different initial disk-surface densities to obtain the lower-density limit for preventing planetary ejection.

\section{Results and Discussion} \label{sec:results}

\begin{figure}
  \centering
  \includegraphics[width=15.0cm]{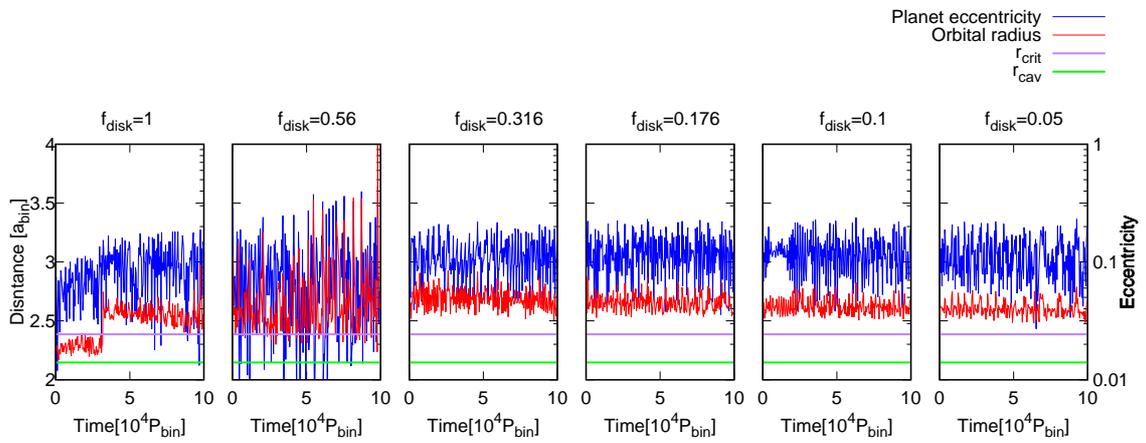}
  \caption{Evolution of the semi-major axis and eccentricity of a planet at each value of $f_{\mathrm{disk}}$ for the $e_{\mathrm{bin}}=0.05, \mu=0.2$ system. Red, blue and purple lines represent the semi-major axis, planet eccentricity, and critical radius respectively. The planet embedded at the inner edge of the circumbinary disk experiences rapid migration to the outside of unstable boundary, followed by modest oscillation near the boundary.\label{fig:e005m02}}
\end{figure}

One result for a system with $e_{\mathrm{bin}}=0.05, \mu=0.2$ is shown in Figure \ref{fig:e005m02}. The planet embedded at the inner edge of the circumbinary disk experiences rapid migration to the outside of unstable boundary when the disk parameter is $f_{\mathrm{disk}}=1$. After escaping the unstable area, the planet's semi-major axis oscillates near the unstable boundary.
In the case that disk of surface density $f_{\mathrm{disk}}=0.56$,
the planet orbit fluctuates from the balancing point, which is very close to the unstable boundary, by the discontinuous change of disk-surface density. This fluctuation causes the planet to repeat entering the unstable area and being pushed back further by the binary gravity, resulting in an increase in the amplitude of oscillation.

In the disk of surface density $f_{\mathrm{disk}}\leq 0.316$, the planet's semi-major axis oscillates just over the unstable boundary with small amplitude.
This is due to the weak negative torque by the weak gas-drag force in these low-density disks and weak orbital excitation because of large planet's orbital distance, which is outside the unstable area.
After the planet moves outside of the unstable region, it sustains a quasi-static orbit over the critical radius. This suggests that, when a planet migrates inward from the outer region, migration would stop before entering the unstable region. Even if a planet entered the unstable region, it would manage to escape to the stable area. The planet's eccentricity stayed around $e_{\mathrm{pl}}\sim 0.1$ throughout this run.

\begin{figure}
  \centering
  \includegraphics[width=15.0cm]{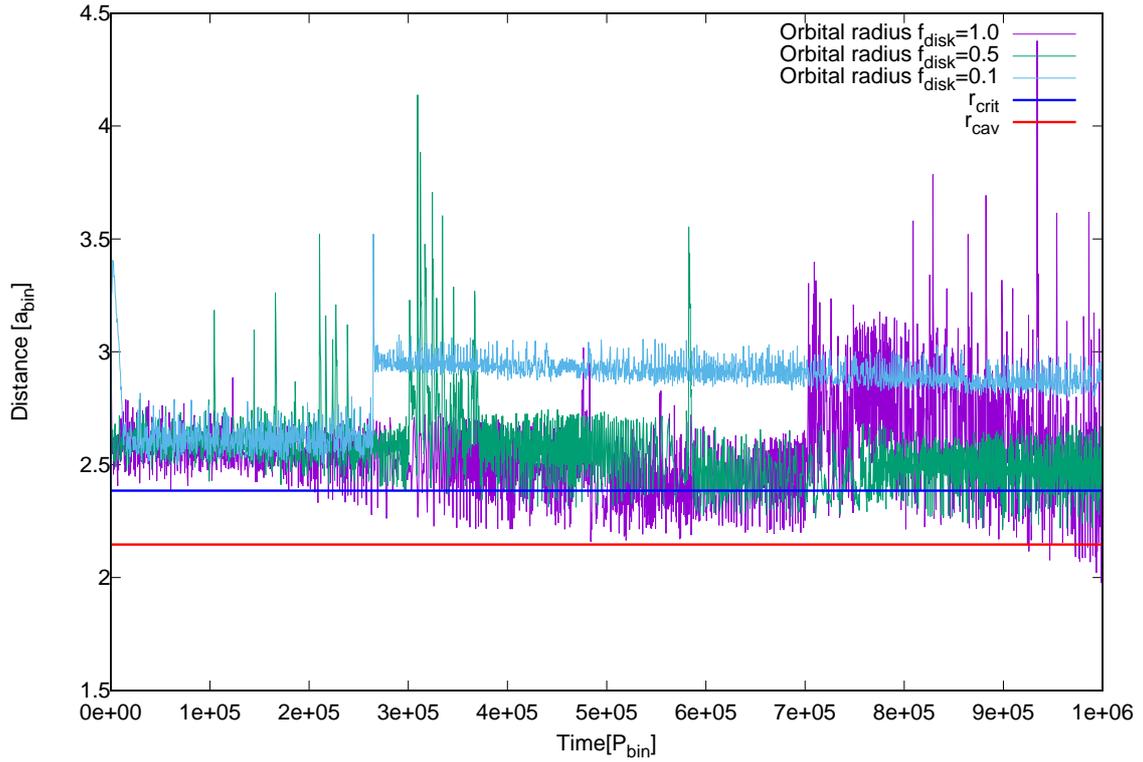}
  \caption{Long-term evolution of the semi-major axis of a planet embedded at $r_{\mathrm{cav}}$ for the $e_{\mathrm{bin}}=0.05, \mu=0.2$ system. Planet orbit expands by accumulation of gravitational instability. Gas-drag avoids strong scattering and, with high $f_{\mathrm{disk}}$, helps to keep the planet orbit close to the instability boundary. }
  \label{fig:longterm_e005m02}
\end{figure}

Figure \ref{fig:longterm_e005m02} shows the orbital evolution of a planet embedded at $r_{\mathrm{cav}}$ in $10^6P_{\mathrm{bin}}$ for $f_{\mathrm{disk}}=1.0, 0.5, 0.1$ to check the long-term stability. The planet's orbit after escaping the unstable area oscillates just over the instability boundary.
We interpret the oscillation is driven by small perturbation of the binary gravity, for the location of the oscillation does not vary with $f_{\mathrm{disk}}$.
The expansion of the planet orbit occurs by gravitational scattering, which is relaxed by the gas-drag to prevent the planet to be ejected. Even outside the unstable area, the planet experiences weak perturbation from the binary and accumulation of this weak perturbations can trigger a sudden instability of the planet orbit. Under a dense gas-disk, negative torque induced by gas-drag damps the planet orbital excitation and the temporarily increased orbital distance is decreased to the original distance. In thin gas-disk with $f_{\mathrm{disk}}=0.1$, the sudden instability transports the planet to another stable orbit at larger distance.

\begin{figure}[p]
  \centering
  \includegraphics[width=15.0cm]{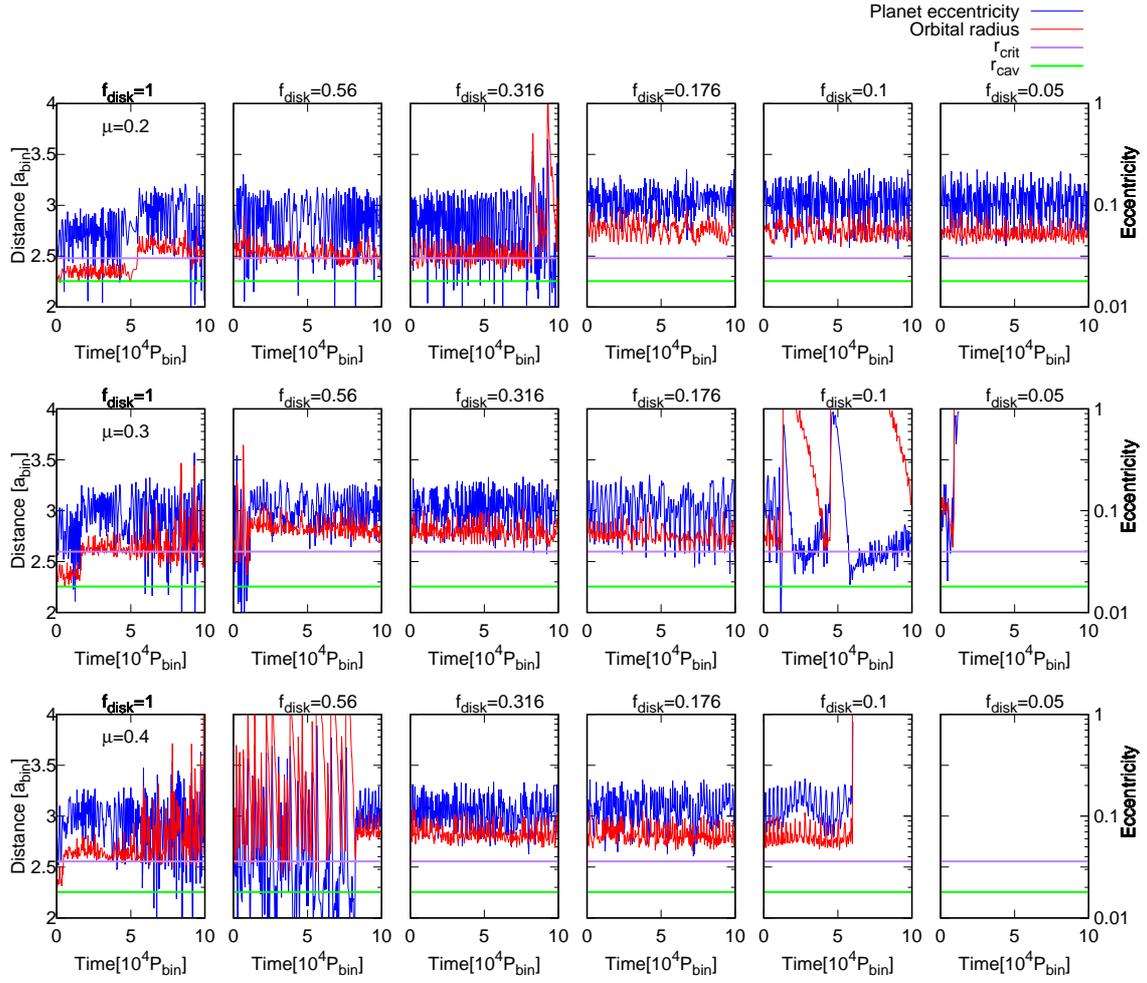}
  \caption{Orbital-evolution of a CBP in $e_{\mathrm{bin}}=0.075$ system. Upper, middle, and lower panels show the results for $\mu=0.2, 0.3, 0.4$. \label{fig:e075}}
\end{figure}

\begin{figure}[p]
  \centering
  \includegraphics[width=15.0cm]{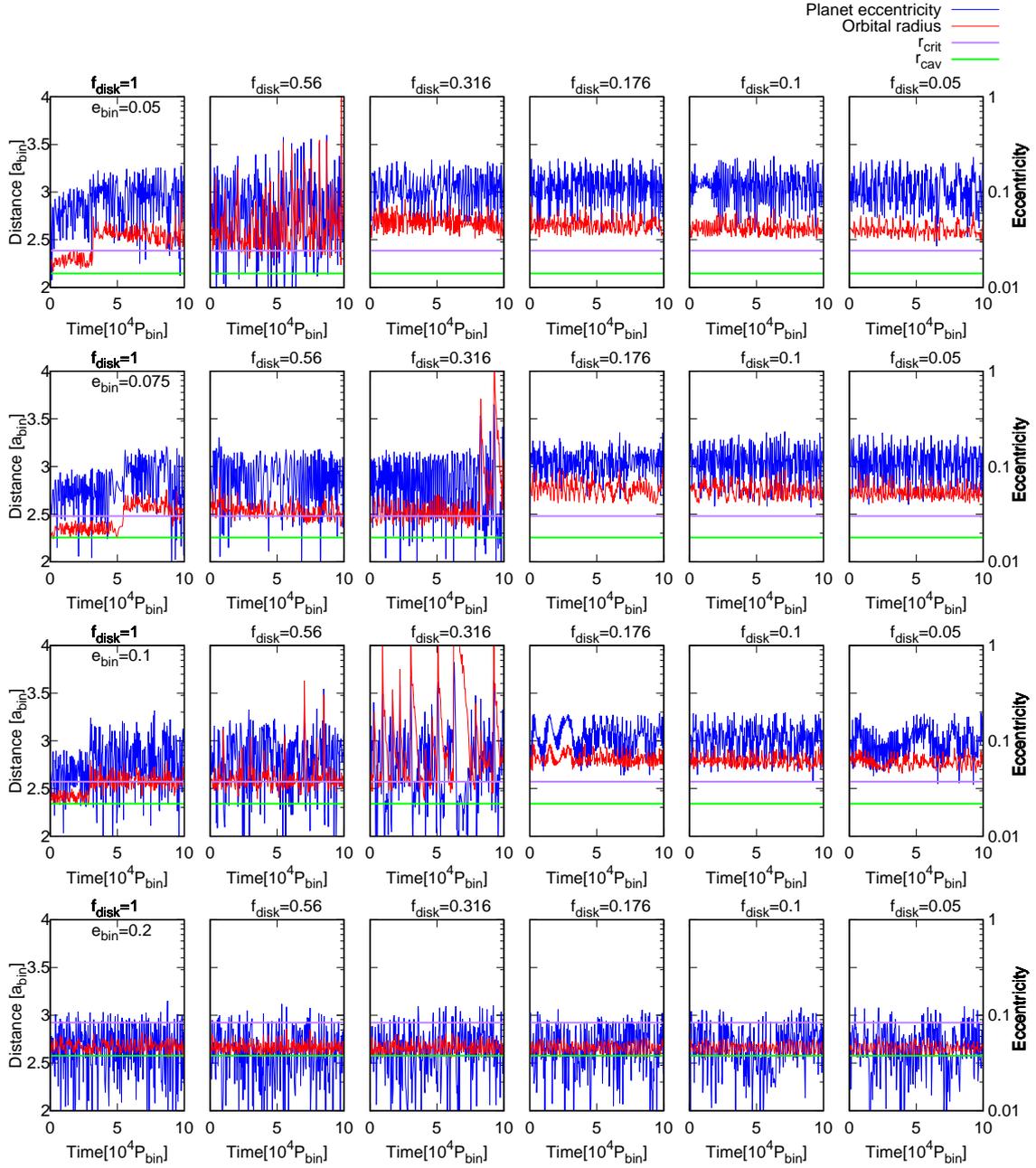}
  \caption{Orbital-evolution of a CBP in $\mu=0.2$ system. Each panel shows the result for $e_{\mathrm{bin}}=0.05, 0.075, 0.1, 0.2$ from top to bottom respectively. \label{fig:m02}}
\end{figure}

Dependencies of the evolution of the semi-major axis and eccentricity of a planet on $\mu$ and $e_{\mathrm{bin}}$ are shown in Figure \ref{fig:e075} and Figure \ref{fig:m02}.
In general, a planet initially embedded at the disk's inner edge moves over the boundary in several thousands of $P_{\mathrm{bin}}$ and stays an orbit just over the unstable boundary when $f_{\mathrm{disk}}=1$ for any set of binary parameters.
The planet first stays at very boundary of the unstable area when it escapes the unstable area. And then, it moves to a larger orbit over the instability boundary after a large oscillation. When the planet is just at the instability boundary, the oscillation amplitude increases when the disk density decreases. And the amplitude decreases as the planet moves to the larger orbit, and then it does not vary when the disk density drops.

Some of the results show a tendency of the average planet's semi-major axis changes on longer timescales as seen in Figure \ref{fig:e005m02}. This would be caused by fluctuation of a planet's semi-major axis due to the discontinuous disk-surface density change. Although the average planet's semi-major axis is stable in each phase of the disk evolution after the planet escaped from the unstable area in general, sometimes it fluctuates when disk-surface density drops discontinuously because the balance of torque from the binary and the disk gas breaks at that moment. The long-time variation of semi-major axis would reflect these fluctuations rather than effects of the disk drag.
In reality, disk-surface density decays smoothly with long timescale. Therefore, the actual planet orbital distance would oscillate just over the instability boundary and at some time, it may experience sudden scattering to a larger orbit as seen in Figure \ref{fig:longterm_e005m02}. When the disk-surface density is high, the planet orbit would shrink to the original distance, where the planet was orbiting before scattering. As the disk-surface density decays, the trend of getting back to the original distance after such orbital jump would decrease and the planet would have a larger orbit. Therefore, CBPs would have orbits just over the instability boundary or little larger.

In low-eccentricity systems with $e_{\mathrm{bin}}=0.05, 0.075,$ oscillation of the semi-major axis is moderate after excited oscillation is damped, especially in low-$\mu$ systems.
Planets in low-eccentricity and low-$\mu$ systems experience weak perturbation of their orbit after reaching the gravitationally stable area and sustain a stable orbit a little outside of the critical radius. Planets experienced perturbations even after they have escaped the unstable region and, in some cases, achieved rather large orbits.

On the other hand, in high-eccentricity systems, the planet is occasionally orbitally excited even after escaping the unstable area, and the planet is ejected from the system.
In systems with $e_{\mathrm{bin}}=0.2$, the gas-drag force is not strong enough to counterbalance the orbital excitation before the disk's surface density drops to 0.1$\Sigma_{\mathrm{MMSN}}$ and the planet escapes the system.
Although in reality a highly eccentric binary should cause non-axisymmetric structures in circumbinary disks, we assume an axisymmetric disk-surface-density profile for simplicity. This approximation may also contribute to the planet ejection.
The $e_{\mathrm{bin}}=0.075, \mu=0.3$ panels of Figure \ref{fig:e075} show that, even if a planet enters the unstable area from outside, it is quickly pushed back to the stable area and stays near the boundary.

If we assume the type-I migration brings the planet to the disk-inner edge, the timescale of the inward migration $\tau_{\mathrm{mig}}$ at the instability boundary is $\tau_{\mathrm{mig}}\sim 6.6 \times 10^3$yr \citep{IdaLIN2008}. Whereas the timescale of outward transportation of the planet to the stable area is $\tau_{\mathrm{transport}} \sim 10^4 P_{\mathrm{bin}} \approx 10^3$yr, which is comparable to $\tau_{\mathrm{mig}}$. Therefore, we suggest that the planet may enter the unstable area by type-I migration, but the planet would be transported outside the unstable boundary by gravitational scattering before reaching the disk's inner edge.
Therefore, our results suggest that even if the unstable area includes the disk's inner edge and a planet migrated to that edge, it could survive orbital evolution by escaping the unstable area and maintaining a stable orbit near the boundary.

\begin{table}
  \centering
  \begin{tabular}{c|ccc}
  \hline
   & $\mu=0.2$ & 0.3 & 0.4 \\
  \hline \hline
  $e_{\mathrm{bin}}=0.05$ & 1 & 0.4 & 0.3\\
  0.075 & 1 & 0.15 & 0.05\\
  0.1 & 0.55 & 0.25 & 0\\
  0.2 & 0.1 & 0 & 0.15\\
  \hline
  \end{tabular}
  \caption{Rate of the case in which the planet can escape the unstable area for systems with different binary parameters. 20 runs were carried out for each setting.}
  \label{tab:para_vs_survival}
\end{table}

The probabilities of a planet managing to escape the unstable area without being ejected from the system among the different sets of 20 runs are shown in Table \ref{tab:para_vs_survival}. Henceforth, we will call this probability the "planet-survival rate." This rate in general decreases as binary eccentricity and mass ratio increase. This suggests that more intense perturbations induced by more non-axisymmetric gravitational fields in eccentric and high $\mu$ binary systems decrease the planet-survival rate.
The planet-survival rate of $e_{\mathrm{bin}}=0.2, \mu=0.4$ is relatively high. This is considered to be due to a statistical reason. More apparent trend that planet-survival rate is higher in lower $e_{\mathrm{bin}}$ and $\mu$ would be seen if more calculations are conducted with every set of $e_{\mathrm{bin}}$ and $\mu$.
In $\mu=0.2$ systems, planets survived migration in all runs in systems less eccentric than $e_{\mathrm{bin}}=0.1$. The survival rate is 0.55 in $e_{\mathrm{bin}}=0.1$ systems and drops to 0.1 in the most eccentric $e_{\mathrm{bin}}=0.2$ systems.
Note that these values have $\sim 5 \%$ uncertainty given 20 runs per set.

The initial disk-surface density in the abovementioned calculations is $f_{\mathrm{disk}}=1$. When the initial disk-surface density is $f_{\mathrm{disk}}\leq0.56$, a planet embedded at the disk's inner edge is quickly ejected, and only $\sim 3 \%$ of the planets achieve stable orbits just over the instability boundary. This is because gas-drag damping is not strong enough to counterbalance the excitation. This corresponds to the case where planetary migration occurs in the later phase of disk evolution. This result indicates that planets are more likely to survive when disk gas is dense, which is consistent with planetary migration being more active at higher gas-surface densities.

\begin{figure}[h]
  \centering
  \includegraphics[width=12.0cm]{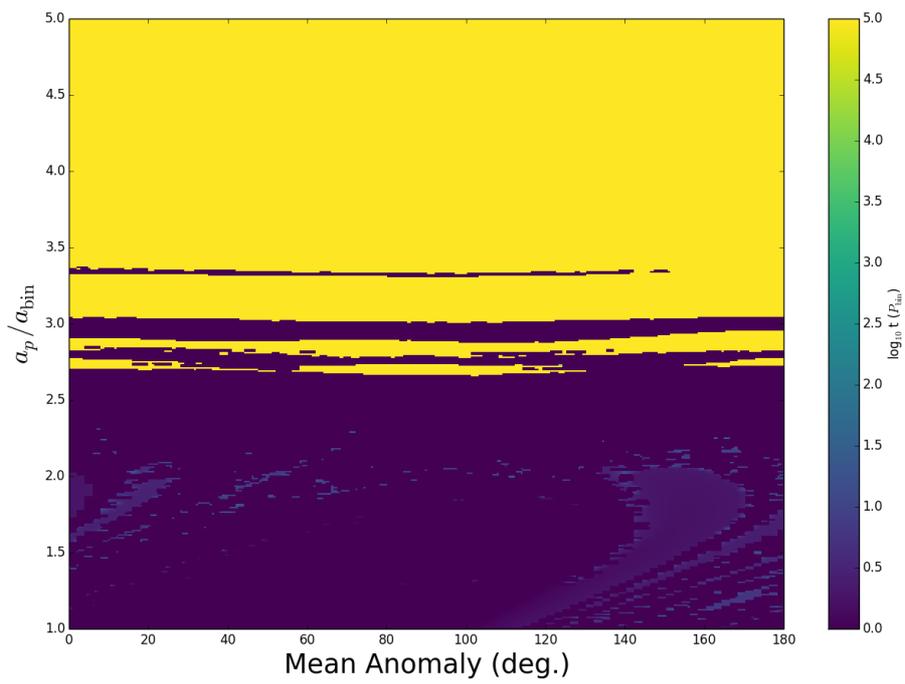}
  \caption{Survival time of a test particle around an $e_{\mathrm{bin}}=0.2, \mu=0.2$ binary system without disk gas calculated using results of \cite{Quarles2018}.
  \label{fig:azimuth}}
\end{figure}

Note that some planets could sustain stable orbits around the initial distance from the binary without being ejected or escaping the unstable area.
Although the planets in such runs were not ejected, we do not consider them as "survivors" in the planet-survival rate (Table \ref{tab:para_vs_survival}). This is because the planets' orbits do not match the characteristics of the observed CBPs.
Using the simulation results of \cite{Quarles2018}, we investigate the effect of initial mean anomaly and orbital distance on the survival time of a test particle (here representing the planet) around a binary star.
For each set of initial mean anomaly and orbital distance, we determined the survival time as the simulation length until the occurrence of instability, which is defined as an intersection with the binary orbit or when the radial distance of the planet from the primary star exceeds 10au in \cite{Quarles2018}.
The dependencies of the survival time on initial mean anomaly and orbital distance are shown in Figure \ref{fig:azimuth}. There are several stable regions in the transitional area between unstable and stable areas.
Both the orbital distance of the planet, which remained unstable area without being destabilized, and the stable region of \cite{Quarles2018} are located between locations of 4:1 and 5:1 mean motion resonances with the binary orbit ($a_{\mathrm{4:1MMR}}\approx 2.52a_{\mathrm{bin}}, a_{\mathrm{5:1MMR}}\approx 2.92a_{\mathrm{bin}}$). These mean motion resonances act to destabilize CBPs \citep{MW2006}. However, it is shown that stability islands exist approximately half-way between the resonances \citep[e.g.][]{DB2011, QL2016}. The planets that remained in the unstable area without being destabilized in our simulations are considered to be trapped in the stability island between 4:1 and 5:1 mean motion resonance.

We speculate the gas drag would not considerably affect the planet survival time for the gas density ranges we adopt in our simulations because the locations where the planets stayed within the unstable area in our simulations with gas drag matches the stable islands which derived from \cite{Quarles2018} without gas drag. But if the gas density is extremely high, it may act to lower the boundary of the stable area.

\section{Comparison with Observations} \label{sec:comparison}

The trend that planets are more likely to survive in less-eccentric binary systems is consistent with the fact that observed binary systems hosting CBPs have low eccentricities of $e_{\mathrm{bin}} \leq 0.2$ (with the exception of Kepler-34b). Binary stars in the Kepler-34 system are highly eccentric ($e_{\mathrm{bin}}=0.521$) with high mass ratio $\mu=0.493$, which indicates a gravitational perturbation is stronger than other observed systems. Strong perturbation makes a large unstable region, and consequently Kepler-34b has a long semi-major axis and high eccentricity. The calculated final eccentricities of the planets are $e_{\mathrm{pl}}\lesssim0.1$, which is comparable to or a little larger than the observations
except for Kepler-34b and Kepler-453b, whose eccentricities are $e_{\mathrm{pl}}=0.182, 0.118$, respectively.

We obtained the planets' final distances from the center of mass from our calculations (Table \ref{tab:fin_a}). Final locations are 2\%-10\% larger than the unstable boundary, except for the higher values in $e_{\mathrm{bin}}=0.2$ systems.
Planets in the $e_{\mathrm{bin}}=0.2$ systems are highly excited, and the calculated final location of these planets are $\sim 40 \%$ larger than the orbital distance range of observed CBPs.
The distances of the observed planets are mostly in the 3.21-3.64 $a_{\mathrm{bin}}$ range. These values are larger than the orbital distance predicted from our calculations in moderate binary eccentricities, $e_{\mathrm{bin}} \leq 0.1$. But, considering that our process of disk dissipation is not continuous, actual planets are expected to experience binary torques for longer time.
As seen in Figure \ref{fig:longterm_e005m02}, planetary orbit can sometimes suddenly expand by accumulation of weak instability even in the stable area, if the planet is exposed to time varying gravity of stellar binary for long time. Since the actual CBPs’ evolutional timescale is longer than our simulation length, the chance of expanding the orbit by instability is expected to be considerable.
Therefore, the actual CBPs would be able to have larger orbits which are consistent with the observed semi-major axes.
Our results suggest that CBPs' orbits just over the instability boundary can be achieved via our transportation scenario even if the planets once entered the unstable area.

\begin{table}
  \centering
  \begin{tabular}{c|ccc}
  \hline
   & $\mu=0.2$ & 0.3 & 0.4 \\
  \hline \hline
  $e_{\mathrm{bin}}=0.05$ & 2.64 & 2.65 & 2.65 \\
  0.075 & 2.75 & 2.73 & 2.85 \\
  0.1 & 2.75 & 2.8 & - \\
  0.2 & 4.06 & - & 4.45 \\
  \hline
\end{tabular}
  \caption{Typical values of final semi-major axis of planets that survived orbital evolution o a stable area relative to binary separation.}
  \label{tab:fin_a}
\end{table}

In previous attempts to reproduce the observed CBPs' orbits via planetary migration from the outer region of the protoplanetary disk, the migration halted at a location which was about 30\%-50\% larger than the disk's surface density peak location (\cite{TK2018}, \cite{PN2013}). On the other hand, in this study, we propose a new scenario: A planet somehow enters the unstable area, and then the planet moves from the unstable area to the outside just over the unstable boundary, which is consistent with the locations of observed CBPs. We suggest that these observed CBPs should have entered the unstable area at a certain stage of disk dissipation when the surface density was high enough to prevent planetary ejection, and then were transported to the current orbits. Since the location of the unstable boundary in a binary system is determined by the separation, eccentricity, and mass ratio of the binary stars, the orbital period of a planet near the boundary is predictable. Therefore, more CBPs are expected to be discovered with periods close to these predictions. Previous studies have shown several options for the mechanism by which a planet falls into inner region, such as shrinkage of the inner cavity of the protoplanetary disk or gravitational interaction with other objects. However, the detailed consideration of this mechanism is left for future work.

In this research we did not consider the non-axisymmetric features of circumbinary disks, which significantly affect highly eccentric systems like Kepler-34, and we assumed a radial profile of the disk-surface density as a power of -1.5, ignoring the non-power-law structure of the density peak (\cite{PN2013}). A more precise surface-density structure is required to conduct this simulation on more eccentric binary systems.

\section{Conclusion} \label{sec:conc}

In our simulations, outward planetary transportation from the disk's inner edge halts just over the unstable boundaries. A planet embedded at the disk's inner edge quickly moves to the stable area by gravitational perturbation from the central binary.
The outward transportation stops when the planet crosses the instability boundary, beyond which perturbations become weak.
After escaping from the unstable area, the planet sustains a stable orbit with its semi-major axis oscillating with moderate amplitude near the boundary, and even if it begins to enter the unstable area again, it gets pushed back outside of the boundary.

The planet-survival rate is high in binary systems with $e_{\mathrm{bin}}\leq0.1, \mu\leq0.3$, which covers the parameters of most of real systems hosting CBPs. Observed CBPs in the migration area near the boundary in our simulations are expected to have experienced the orbital evolution considered here. The planet is ejected from the system when the initial disk-surface density is under $0.56\Sigma_{\mathrm{MMSN}}$, indicating that a planet cannot escape the unstable area without being ejected unless there is enough disk gas, which is also a favorable condition for planet migration.

We conclude that CBPs have to enter the unstable area when enough amount of gaseous disk remains to maintain the CBPs' stable orbits just over the instability boundary.
Our results suggest that, when a migrating planet crosses the unstable boundary it gets pushed back to stable area where orbital excitation is weak and starts to move inward again due to gas-drag damping. Close to the unstable boundary, this orbit is achievable by repeating this process. Still, since previous studies have shown that inward planetary migration halts at a disk-density peak that is more distant than observed planet locations, mechanisms for shrinking the disk inner cavity which enables a planet to migrate inside or close to the unstable area are needed.
In this study, we considered an axisymmetric disk-surface-density profile with a radial structure of power law of -1.5 for simplicity. However, a more precise circumbinary-disk, surface-density profile including non-axisymmetric features and a density bump near the inner cavity may be used to apply our concept to highly eccentric binary systems. Checking the long-term stability of large planets also remains as future work.

\section{List of Abbreviation}

\begin{itemize}
  \item CBPs: Circumbinary planets
  \item MMSN: Minimum mass solar nebula
  \item MMR:  Mean motion resonance
\end{itemize}

\section{Declarations}

\subsection{Availability of data and materials}

The datasets used and/or analysed during the current study are available from the corresponding author on reasonable request.

\subsection{Competing interests}

The authors declare that they have no competing interests.

\subsection{Funding}

This work was supported by MEXT Grant Number 26106006 and KURA Research Development Program ISHIZUE.

\subsection{Author's contributions}

AY designed the study, analyzed the data and wrote the manuscript. TS contributed to improvement of the discussion and revision of the draft manuscript. Both of the authors read and approved the final manuscript.

\subsection{Acknowledgements}

We thank anonymous reviewers for their constructive comments, which led us to greatly improve this paper.


\bibliographystyle{bmc-mathphys} 
\bibliography{eps_bib}      

\nocite{label}

\end{document}